# Free-standing ferroelectric multilayers: Crossover from thin-film to bulk behavior

S. Prokhorenko[1] and N. A. Pertsev[2]

[1]*St. Petersburg Academic University–Nanotechnology Research and Education Centre of the Russian Academy of Sciences, 194021 St. Petersburg, Russia*

[2]*A. F. Ioffe Physico-Technical Institute, Russian Academy of Sciences, 194021 St. Petersburg, Russia*

Ferroelectric films usually have phase states and physical properties very different from those of bulk ferroelectrics. Here we propose free-standing ferroelectric-elastic multilayers as a bridge between these two material systems. Using a nonlinear thermodynamic theory, we determine phase states of such multilayers as a function of temperature, misfit strain, and volume fraction $\phi_f$ of ferroelectric material. The numerical calculations performed for two classical ferroelectrics − PbTiO$_3$ and BaTiO$_3$ − demonstrate that polarization states of multilayers in the limiting cases $\phi_f \rightarrow 0$ and $\phi_f \rightarrow 1$ coincide with those of thin films and bulk crystals. At intermediate volume fractions, however, the misfit strain-temperature phase diagrams of multilayers differ greatly from those of epitaxial films. Remarkably, a ferroelectric phase not existing in thin films and bulk crystals can be stabilized in BaTiO$_3$ multilayers. Owing to additional tunable parameter and reduced clamping, ferroelectric multilayers may be superior for a wide range of device applications.

## I. INTRODUCTION

Ferroelectrics represent an important class of functional materials, with a wide range of applications in electronic devices exploiting their unique polarization, dielectric, and piezoelectric properties.[1] Remarkably, phase states and physical properties of ferroelectrics usually strongly differ in thin-film and bulk forms. As ferroelectric thin films are predominantly fabricated on dissimilar thick substrates, this difference largely results from the substrate-induced straining and clamping of a thin film.[2] The most impressive manifestations of this mechanical film-substrate interaction are the appearance of new ferroelectric phases forbidden in bulk crystals[2-5] and the phenomenon of strain-induced ferroelectricity in films of incipient ferroelectrics such as SrTiO$_3$ and KTaO$_3$.[6,7,8] In ultrathin films, the depolarizing-field effect and the short-range interactions at film surfaces also come into play, leading to the scaling of spontaneous polarization, ferroelectric transition temperature, coercive field, and other physical characteristics.[9-16] The influence of these effects on the film overall performance, however, gradually becomes less and less pronounced as the film thickness $t$ increases so that they may be neglected for films sandwiched between metallic electrodes already at a thickness $t$ ~100 nm. In contrast, the substrate-induced lattice strains remain to be considerable in this thickness range,[17-19] suitable for nanoelectronic devices, and the substrate clamping effect cannot be ignored until the film becomes much thicker than the substrate.

In order to reveal the full potential of strain engineering for ferroelectric materials, it is important to describe the *crossover* from thin-film to bulk behavior. In this paper, we propose free-standing multilayers comprising ferroelectric slabs interleaved by non-ferroelectric layers with linear elastic properties as a "bridge" material system displaying such crossover behavior. It is shown that the mechanical boundary conditions imposed on the ferroelectric component in such multilayers are *intermediate* between the two-dimensional clamping characteristic of a thin film and the stress-free state of a bulk crystal. To avoid bending effects, which are absent in free-standing crystals and negligible in films deposited on thick substrates, we consider only multilayers that are symmetric with respect to the central slab, involving odd number of ferroelectric and even number of passive layers (Fig. 1). The absence of substrate-induced clamping and straining makes the proposed material system very different from conventional ferroelectric superlattices and mul-

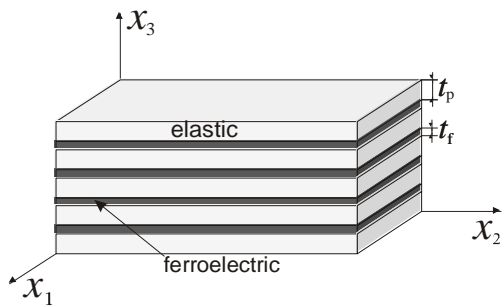

**FIG. 1**. Schematic representation of a free-standing multilayer involving odd number of ferroelectric and even number of elastic (passive) slabs.

tilayers, which were studied in many experimental and theoretical works.[20-29]

For multilayers with single-domain ferroelectric slabs, we develop a nonlinear thermodynamic model enabling us to determine their equilibrium polarization states as a function of temperature, internal strains caused by lattice mismatch between constituents, and volume fraction $\phi_f$ of ferroelectric material (Sec. II). The strain-temperature phase diagrams of $PbTiO_3$ and $BaTiO_3$ multilayers are constructed and compared with known thin-film diagrams (Sec. III). It should be emphasized that our model is also applicable to *ferroelectric-ferromagnetic* multilayers, as it describes their ground polarization state in the absence of external magnetic fields. Therefore, our results provide the basis for theoretical calculations of a strain-mediated magnetoelectric effect in such multiferroic multilayers, which should be enhanced due to the absence of a thick passive substrate.

## II. THERMODYNAMIC MODEL

Our approach is based on the nonlinear thermodynamic theory of ferroelectric crystals[30,31] and thin films.[2] The Helmholtz free energy density $F_f$ of a ferroelectric material is described by a polynomial in terms of three polarization components $P_i$ ($i$ = 1,2,3). This polynomial also takes into account the coupling between polarization and strain, which results from the electrostriction. The non-ferroelectric component of the multilayer is treated as a linear elastic material, which may be conductive or insulating. Therefore, the energy density $F_p$ of "passive" layers generally contains contributions from both the elastic strain energy and the electrostatic energy, which is associated with internal electric fields created by the polarization charges existing at the surfaces of ferroelectric slabs.[24,27] Since this paper is devoted to the mechanical aspect of the problem, we will focus on the case of multilayers, where the electrostatic energy of passive layers and the depolarizing-field energy of ferroelectric ones may be neglected. This case corresponds predominantly to metallic passive layers providing almost perfect screening of interfacial polarization charges by free charge carriers. In the end of this section, we will show how the electrostatic energy can be taken into account in the case of insulating passive layers.

Restricting our analysis to single-domain states of ferroelectric slabs, we can assume the polarization to be uniform inside each slab[32] and so replace the total free energy of the multilayer by the mean energy density

$$\langle F \rangle = \phi_f F_f + \phi_p F_p, \qquad (1)$$

where $\phi_f = t_f/(t_f + t_p)$ and $\phi_p = t_p/(t_f + t_p)$ are the volume fractions of ferroelectric and passive materials in the multilayer, which are determined by the thicknesses $t_f$ and $t_p$ of individual layers. Evidently, $\langle F \rangle$ depends on the polarization components $P_i$ and on the lattice strains in ferroelectric and passive slabs, which will be denoted by $u_{ij}$ and $v_{ij}$, respectively. To determine the equilibrium polarization state of ferroelectric layers via the minimization of $\langle F \rangle$, we must reduce Eq. (1) to a function of polarization components and misfit strains in the multilayer only. This transformation can be performed using the conditions of mechanical equilibrium and strain compatibility at the interfaces.

To that end, we first take into account that there are no external mechanical forces acting on the outer surface of a free-standing multilayer. Considering this boundary condition for the surface sections parallel to the interfaces (i.e., orthogonal to the $x_3$ axis of our coordinate system shown in Fig. 1), we find that the stresses $\sigma_{i3}$ ($i = 1,2,3$) are absent in the whole multilayer. Hence $\partial F_f / \partial u_{i3} = 0$ and $\partial F_p / \partial v_{i3} = 0$ so that the strains $u_{i3}$ and $v_{i3}$ can be expressed through the in-plane strains $u_{\alpha\beta}$ and $v_{\alpha\beta}$ ($\alpha,\beta = 1,2$) in ferroelectric and passive layers. Next, the calculation of forces acting on the surface sections perpendicular to the $x_1$ and $x_2$ axes gives $\phi_f \sigma_{\alpha\beta} + \phi_p \sigma^p_{\alpha\beta} = 0$, where $\sigma_{\alpha\beta}$ and $\sigma^p_{\alpha\beta}$ denote the stresses in ferroelectric and passive layers, respectively. Since $\sigma_{\alpha\beta} = \partial F_f / \partial u_{\alpha\beta}$ and $\sigma^p_{\alpha\beta} = \partial F_p / \partial v_{\alpha\beta}$, these equations can be rewritten as

$$\varphi_f \frac{\partial F_f}{\partial u_{11}} + \varphi_p \frac{\partial F_p}{\partial v_{11}} = 0, \quad \varphi_f \frac{\partial F_f}{\partial u_{12}} + \varphi_p \frac{\partial F_p}{\partial v_{12}} = 0,$$

$$\varphi_f \frac{\partial F_f}{\partial u_{22}} + \varphi_p \frac{\partial F_p}{\partial v_{22}} = 0, \qquad (2)$$

thus giving three relations between $u_{\alpha\beta}$ and $v_{\alpha\beta}$. The remaining three relationships follow from the condition of strain compatibility at the interfaces, which yields

$$u_{11} = v_{11} + u_{m1}, \quad u_{22} = v_{22} + u_{m2},$$

$$u_{12} = v_{12} + u_{m6}, \qquad (3)$$

where $u_{m1}$, $u_{m2}$, and $u_{m6}$ are the misfit strains in the multilayer. In the case of (001)-oriented layers of ferroelectric materials having cubic paraelectric phase, these strains can be calculated as $u_{m1} = (b_1 - a_0)/a_0$, $u_{m2} = (b_2 - a_0)/a_0$, and $u_{m6} = \gamma - \pi/2$, where $a_0$ is the equivalent cubic cell constant of a stress-free ferroelectric crystal, $b_1$ and $b_2$ are the lattice parameters of free-standing passive layers measured along the $x_1$ and $x_2$ axes, whilst $\gamma$ is the angle between their crystallographic directions in the $x_1x_2$ plane related to the lattice matching with ferroelectric layers. The system of six equations (2)-(3) makes it possible to calculate the in-plane strains $u_{\alpha\beta}$ and $v_{\alpha\beta}$ as functions of three misfit strains and three polarization components. The substitution of resulting relations into Eq. (1) yields the sought expression for the mean energy density $\langle F \rangle$. In this work, we performed necessary calculations for multilayers involving (001)-oriented slabs of a perovskite ferroelectric and a passive material with cubic symmetry. In this practically important case, $u_{m1} = u_{m2} = u_m$, $u_{m6} = 0$, and the strain energy $F_p$ reduces to

$$F_p = \frac{1}{2}c_{11}^p\left(v_{11}^2 + v_{22}^2 + v_{33}^2\right) + c_{12}^p\left(v_{11}v_{22} + v_{11}v_{33} + v_{22}v_{33}\right)$$

$$+ \frac{1}{2}c_{44}^p\left(v_{12}^2 + v_{13}^2 + v_{23}^2\right), \quad (4)$$

where $c_{ln}^p$ are the elastic stiffnesses of passive material in its crystallographic reference frame. The ferroelectric energy density $F_f$ was derived from the Gibbs free energy $G$ via the inverse Legendre transformation $F_f = G + \sum_{j \geq i} u_{ij}\sigma_{ij}$, which enables us to use the thermodynamic coefficients determined for bulk crystals.[31,33] As we consider ferroelectrics with a cubic prototypic state, the basic expression for the Gibbs free energy has the form

$$G = \alpha_1(P_1^2 + P_2^2 + P_3^2) + \alpha_{11}(P_1^4 + P_2^4 + P_3^4) + \alpha_{12}(P_1^2P_2^2 + P_1^2P_3^2 + P_2^2P_3^2) + \ldots - Q_{11}(\sigma_{11}P_1^2 + \sigma_{22}P_2^2 + \sigma_{33}P_3^2) - Q_{12}\left[\sigma_{11}\left(P_2^2 + P_3^2\right) + \sigma_{22}\left(P_1^2 + P_3^2\right) + \sigma_{33}\left(P_1^2 + P_2^2\right)\right] - Q_{44}(\sigma_{12}P_1P_2 + \sigma_{13}P_1P_3 + \sigma_{23}P_2P_3)$$
$$- \frac{1}{2}s_{11}(\sigma_{11}^2 + \sigma_{22}^2 + \sigma_{33}^2) - s_{12}(\sigma_{11}\sigma_{22} + \sigma_{11}\sigma_{33} + \sigma_{22}\sigma_{33}) - \frac{1}{2}s_{44}(\sigma_{12}^2 + \sigma_{13}^2 + \sigma_{23}^2), \quad (5)$$

where $\alpha_1$ and $\alpha_{ij}$ represent the dielectric stiffness and higher-order stiffness coefficients at constant stress, $s_{ln}$ are the elastic compliances at constant polarization, and $Q_{ln}$ are the electrostrictive constants of paraelectric phase. Setting $\sigma_{13} = \sigma_{23} = \sigma_{33} = 0$ and using the relations $u_{\alpha\beta} = -\partial G/\partial \sigma_{\alpha\beta}$ to express other stresses through the in-plane strains, one can derive from Eq. (5) the energy $F_f(P_i, u_{\alpha\beta})$ needed to calculate the mean energy density $\langle F \rangle(P_i, u_m)$. After some algebra, the procedure outlined above yields

$$\langle F \rangle = \phi_f \cdot [\alpha_1^*(P_1^2 + P_2^2) + \alpha_3^*P_3^2 + \alpha_{11}^*(P_1^4 + P_2^4) + \alpha_{33}^*P_3^4 + \alpha_{12}^*P_1^2P_2^2 + \alpha_{13}^*(P_1^2P_3^2 + P_2^2P_3^2) +$$
$$+ \alpha_{111}(P_1^6 + P_2^6 + P_3^6) + \alpha_{112}(P_1^2(P_2^4 + P_3^4) + P_2^2(P_1^4 + P_3^4) + P_3^2(P_1^4 + P_2^4)) + \alpha_{123}P_1^2P_2^2P_3^2 +$$
$$+ \alpha_{1111}(P_1^8 + P_2^8 + P_3^8) + \alpha_{1112}(P_1^6(P_2^2 + P_3^2) + P_2^6(P_1^2 + P_3^2) + P_3^6(P_1^2 + P_2^2)) +$$
$$+ \alpha_{1122}(P_1^4P_2^4 + P_2^4P_3^4 + P_1^4P_3^4) + \alpha_{1123}(P_1^4P_2^2P_3^2 + P_2^4P_1^2P_3^2 + P_3^4P_1^2P_2^2)]. \quad (6)$$

Here $\alpha_i^*$ and $\alpha_{ij}^*$ denote the renormalized thermodynamic coefficients, and the polarization terms up to eighth order are included to allow the description of BaTiO$_3$,[33] whilst terms independent of polarization are omitted. Equation (6) demonstrates that the mechanical interaction with passive layers renormalizes only the second- and fourth-order polarization terms. The sixth- and higher-order terms appear to be insensitive to mechanical boundary conditions and, therefore, are the same for bulk crystals, thin films, and multilayers. The renormalized second-order coefficients $\alpha_i^*$ involved in Eq. (6) are given by

$$\alpha_1^* = \alpha_1 - \frac{(Q_{11}+Q_{12})\phi_p u_m}{(s_{11}^p+s_{12}^p)(1-\phi_p)+(s_{11}+s_{12})\phi_p},$$

$$\alpha_3^* = \alpha_1 - \frac{2Q_{12}\phi_p u_m}{(s_{11}^p+s_{12}^p)(1-\phi_p)+(s_{11}+s_{12})\phi_p}, \quad (7)$$

where $s_{ln}^p$ are the elastic compliances of passive layers. It can be seen that, in similarity with the case of an epitaxial thin film,[2] $\alpha_i^*$ linearly depend on the misfit strain $u_m$. At the same time, the renormalized fourth-order coefficients $\alpha_{ij}^*$ are independent of $u_m$ and can be written as

$$\alpha_{11}^* = \alpha_{11} + \frac{(Q_{11}^2+Q_{12}^2)s_{11}-2Q_{11}Q_{12}s_{12}}{2(s_{11}^2-s_{12}^2)} - \frac{(Q_{11}+Q_{12})^2(s_{11}^p+s_{12}^p)(1-\phi_p)}{2(s_{11}+s_{12})[(s_{11}^p+s_{12}^p)(1-\phi_p)+(s_{11}+s_{12})\phi_p]} - \frac{1}{2}\beta(\phi_p),$$

$$\alpha_{12}^* = \alpha_{12} + \frac{2Q_{11}Q_{12}s_{11}-(Q_{11}^2+Q_{12}^2)s_{12}}{s_{11}^2-s_{12}^2} + \frac{Q_{44}^2\phi_p}{2[s_{44}^p(1-\phi_p)+s_{44}\phi_p]} + \beta(\phi_p), \quad (8)$$

$$\alpha_{13}^* = \alpha_{12} + \frac{2(Q_{11}+Q_{12})Q_{12}\phi_p}{(s_{11}^p+s_{12}^p)(1-\phi_p)+(s_{11}+s_{12})\phi_p}, \quad \alpha_{33}^* = \alpha_{11} + \frac{Q_{12}^2\phi_p}{(s_{11}^p+s_{12}^p)(1-\phi_p)+(s_{11}+s_{12})\phi_p},$$

$$\beta(\phi_p) = \frac{(Q_{11}-Q_{12})^2(s_{11}^p-s_{12}^p)(1-\phi_p)}{2(s_{11}-s_{12})[(s_{11}^p-s_{12}^p)(1-\phi_p)+(s_{11}-s_{12})\phi_p]} - \frac{(Q_{11}+Q_{12})^2(s_{11}^p+s_{12}^p)(1-\phi_p)}{2(s_{11}+s_{12})[(s_{11}^p+s_{12}^p)(1-\phi_p)+(s_{11}+s_{12})\phi_p]},$$

where $\beta(\phi_p)$ goes to zero at $\phi_p = 1$. The analysis of Eqs. (7) and (8) demonstrates that in the limiting case of $\phi_p = 1$ our relations for the renormalized second- and fourth-order coefficients transform into those derived earlier for an epitaxial ferroelectric film deposited on a thick substrate.[2] At $\phi_p = 0$, the coefficients $\alpha_i^*$ and $\alpha_{ij}^*$ reduce to the dielectric stiffnesses $\alpha_1$ and $\alpha_{ij}$ of the bulk crystal. These results support the validity of our calculations.

Finally, let us determine the electrostatic contribution to the mean free energy $\langle F \rangle$, which appears in the case of multilayers involving insulating passive slabs. Assuming that the multilayer is sandwiched between two short-circuited electrodes, we may write this contribution as

$$\Delta\langle F\rangle = \phi_f\left(\frac{1}{2}\varepsilon_0 E_f^2\right) + \phi_p\left(\frac{1}{2}\varepsilon_0\varepsilon_p E_p^2\right), \quad (9)$$

where $E_f$ and $E_p$ are the internal electric fields in ferroelectric and passive layers, respectively, $\varepsilon_p$ is the relative permittivity of passive layers, and $\varepsilon_0$ is the permittivity of the vacuum. For the fields $E_f$ and $E_p$, the continuity of the electric displacement $\mathbf{D} = \varepsilon_0 \mathbf{E} + \mathbf{P}$ across the interfaces and the condition of zero mean electric field in the multilayer give

$$E_f = -\frac{\phi_p P_3}{\varepsilon_0 [\phi_p + \varepsilon_p (1-\phi_p)]},$$

$$E_p = -\frac{(1-\phi_p) P_3}{\varepsilon_0 [\phi_p + \varepsilon_p (1-\phi_p)]}. \quad (10)$$

The substitution of these relations into Eq. (9) shows that the electrostatic contribution simply renormalizes the second-order polarization term $\alpha_3^* P_3^2$ in Eq. (6) for the multilayer energy density $\langle F \rangle$. The renormalized coefficient $\alpha_3^*$ is defined by the formula

$$\alpha_3^* = \alpha_1 - \frac{2 Q_{12} \phi_p u_m}{(s_{11}^p + s_{12}^p)(1-\phi_p) + (s_{11}+s_{12})\phi_p}$$

$$+ \frac{\phi_p}{2\varepsilon_0 [\phi_p + \varepsilon_p (1-\phi_p)]}. \quad (11)$$

Since the dielectric stiffness $\alpha_1$ is a linear function of temperature $T$ whilst all other involved material parameters may be regarded as temperature-independent in our case, the influence of internal electric fields on the multilayer polarization state is equivalent to a simultaneous change of the misfit strain and temperature.[27] Using the procedure described in Ref. 27, for the effective temperature and misfit strain we obtain

$$\tilde{T} = T + \frac{(Q_{11}+Q_{12})}{(Q_{11}-Q_{12})} \frac{C \phi_p}{[\phi_p + \varepsilon_p (1-\phi_p)]},$$

$$\tilde{u}_m = u_m + \frac{(s_{11}^p + s_{12}^p)(1-\phi_p) + (s_{11}+s_{12})\phi_p}{2\varepsilon_0 (Q_{11}-Q_{12})[\phi_p + \varepsilon_p (1-\phi_p)]}, \quad (12)$$

where $C$ is the Curie-Weiss constant of the ferroelectric material. Hence the multilayer polarization state, which exists in the presence of electrostatic effect at the temperature $T$ and misfit strain $u_m$, can be found by calculating the state appearing in the absence of internal electric fields at the temperature $\tilde{T}$ and strain $\tilde{u}_m$ given by Eq. (12).

Therefore, in the following section we present only the results obtained under the assumption of negligible electrostatic effect.

### III. PHASE DIAGRAMS OF FERROELECTRIC MULTILAYERS

Minimization of the mean free energy $\langle F \rangle$ as a function of polarization components $P_i$ makes it possible to determine the energetically most favorable polarization state of single-domain ferroelectric layers. In general, this state depends on three misfit strains, $u_{m1}$, $u_{m2}$, and $u_{m6}$, temperature $T$, and volume fraction $\phi_p$ of passive material in the multilayer. In our case of isotropic biaxial misfit strain ($u_{m1} = u_{m2} = u_m$, $u_{m6} = 0$), the set of parameters determining the polarization state reduces to three quantities: $u_m$, $T$, and $\phi_p$. Although the misfit strain is governed by the lattice constants of multilayer constituents and weakly changes with temperature,[4] we shall consider $u_m$ as an independent variable as well. This approach is partly justified by the fact that the misfit strain at the growth temperature can be tuned between the nominal value $u_m = (b-a_0)/a_0$ and zero by changing the layer thicknesses.[19] Moreover, it enables us to reveal general features of the polarization behavior of BaTiO$_3$ and PbTiO$_3$ multilayers involving various passive materials. To that end, we will present our results in the form of the *volume fraction-temperature* and *misfit strain-temperature* phase diagrams showing the stability ranges of possible polarization states at a fixed misfit strain and a fixed volume fraction of passive material, respectively.

The numerical calculations of polarization components and phase maps were performed for PbTiO$_3$ and BaTiO$_3$ multilayers using Eqs. (6)-(8) with the thermodynamic coefficients taken from Refs. 31 and 33, respectively. For the electrostrictive constants and elastic compliances of PbTiO$_3$ and BaTiO$_3$, we employed the values listed in Ref. 2. The elastic constants of non-ferroelectric layers were set equal to those of CoFe$_2$O$_4$ ($c_{11}^p = 286$ GPa, $c_{12}^p = 173$ GPa, $c_{44}^p = 45$ GPa, see Ref. 34) treated as a model passive material. Since the elastic stiffnesses of many suitable non-ferroelectric materials have the same order of magnitude as the stiffnesses of CoFe$_2$O$_4$, the developed phase diagrams can be used for qualitative predictions of the polarization states forming in PbTiO$_3$ and BaTiO$_3$ multilayers involving other passive materials.

Let us discuss first the phase maps of PbTiO$_3$ multilayers, which are simpler than BaTiO$_3$ diagrams because PbTiO$_3$ bulk crystals have only one stable ferroelectric phase.[31] Figure 2 shows the evolution of the volume fraction-temperature map with the variation of misfit strain $u_m$ from zero to $-3\times10^{-3}$. In similarity with the bulk PbTiO$_3$, the multilayer has only two stable phases (paraelectric and ferroelectric) at $u_m \leq 0$. However, the mechanical interaction with passive layers lowers the symmetry of the paraelectric phase from cubic to tetragonal. The ferroelectric $c$ phase is also tetragonal and has a polarization orthogonal to the interfaces ($P_1 = P_2 = 0$, $P_3 \neq 0$). Depending on the misfit strain, the temperature $T_{p\to c}$ of ferroelectric phase transition either decreases or increases with the volume fraction $\phi_p$ of passive material (or even has a shallow minimum, see Fig. 2). Remarkably, at a critical value $\phi_p^*$, the first-order phase transition characteristic of bulk PbTiO$_3$ transforms into the second-order transition. Thus, the multilayer with $\phi_p = \phi_p^*$ is distinguished by a *tricritical point*. The volume fraction $\phi_p^*$ is defined by the equality $a_{33}^* = 0$ which gives

$$\phi_p^* = \frac{\alpha_{11}(s_{11}^p + s_{12}^p)}{\alpha_{11}(s_{11}^p + s_{12}^p - s_{11} - s_{12}) - Q_{12}^2}. \quad (13)$$

Equation (13) demonstrates that $\phi_p^*$ is independent of the misfit strain and equals about 0.512 for the discussed PbTiO$_3$ multilayers. However, this relation holds only in the strain range $u_m \leq 0$, where the ferroelectric transition results in the formation of the $c$ phase. The temperature $T_{p\to c}$ of this transition at $\phi_p \geq \phi_p^*$ can be calculated analytically from the equation $a_3^* = 0$ which yields

$$T_{p\to c} = \theta + \frac{4\varepsilon_0 C Q_{12} \phi_p u_m}{(s_{11}^p + s_{12}^p)(1-\phi_p) + (s_{11} + s_{12})\phi_p}, \quad (14)$$

where $\theta$ is the Curie-Weiss temperature. Equation (14) shows that $T_{p\to c}$ linearly depends on the misfit strain, being above $\theta$ at $u_m < 0$ since $Q_{12}$ is negative for PbTiO$_3$.

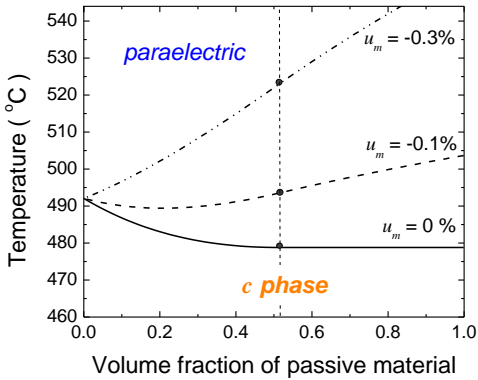

**FIG. 2.** Evolution of the volume fraction-temperature phase map of PbTiO$_3$ multilayer in the strain range $u_m \leq 0$. The transition lines separating stability ranges of the paraelectric phase and the ferroelectric $c$ phase are plotted for three different strain values: $u_m = 0$, $-1\times10^{-3}$, and $-3\times10^{-3}$. The tricritical points on the transition lines are indicated by the dots, whilst the corresponding critical volume fraction is shown by the dashed line.

The structure of ($\phi_p$, $T$) maps becomes more complex for multilayers with positive misfit strains. As can be seen from the representative map shown in Fig. 3, four different ferroelectric states form in this case. In addition to the $c$ phase, the set of stable states includes the orthorhombic $a$ ($P_1 \neq 0$, $P_2 = P_3 = 0$) and $aa$ ($|P_1| = |P_2| \neq 0$, $P_3 = 0$) phases with polarizations parallel to the interfaces and the monoclinic $r$ phase with the polarization inclined to them ($|P_1| = |P_2| \neq 0$, $P_3 \neq 0$). The ($\phi_p$, $T$) diagram is distinguished by the presence of triple points, where three different phases meet. The most interesting triple point lies on the line $T_c(\phi_p)$ of the ferroelectric phase transition, being located at a certain volume fraction $\phi_p^0$ of passive material. Indeed, when $\phi_p$ exceeds $\phi_p^0$, the computed misfit strain-temperature phase map of the multilayer changes qualitatively and becomes similar to the ($u_m$, $T$) diagram of a ferroelectric thin film. On the contrary, the stability range of the $a$ phase appears on the multilayer map at $\phi_p < \phi_p^0$ (see Fig. 4), which is absent in the phase diagram of PbTiO$_3$ film.[2] The threshold volume fraction $\phi_p^0$ is independent of the misfit strain and can be estimated from the equation $\alpha_{11}^*(\phi_0) = \alpha_{12}^*(\phi_0)/2$, which gives $\phi_p^0 \cong 0.815$ for our PbTiO$_3$ multilayers.

At positive misfit strains, the tricritical point also appears at a certain volume fraction $\phi_p^*$. However, the magnitude of $\phi_p^*$ changes from 0.512 at $u_m \to 0$ to about 0.12 at $u_m \cong 0.817 \cdot 10^{-3}$. Remarkably, within this strain range the tricritical point coincides with the triple point, at which the paraelectric, $c$, and $a$ phases meet. At larger positive strains, the tricritical point corresponds to the transition into $a$ phase so that $\phi_p^*$ stabilizes at the value of 0.12 being the root of the equation $a_{11}^* = 0$. The temperature $T_{p \to a}$ of this phase transition at $\phi_p \geq \phi_p^*$ can be calculated analytically from the equation $a_1^* = 0$ and is given by the formula

$$T_{p \to a} = \theta + \frac{2\varepsilon_0 C(Q_{11} + Q_{12})\phi_p u_m}{(s_{11}^p + s_{12}^p)(1 - \phi_p) + (s_{11} + s_{12})\phi_p}. \quad (15)$$

We see that $T_{p \to a}$ linearly increases with the strain $u_m > 0$ since $Q_{11} + Q_{12}$ is positive for PbTiO$_3$.

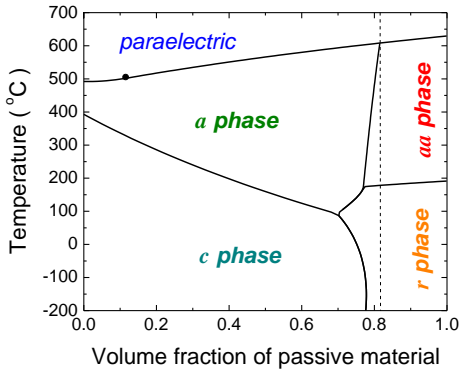

**FIG. 3.** Volume fraction-temperature phase map of PbTiO$_3$ multilayer with $u_m = +5 \times 10^{-3}$. The dashed line shows the threshold volume fraction $\phi_p^0$, above which the multilayer ($u_m$, $T$) diagram becomes similar to that of the PbTiO$_3$ film.

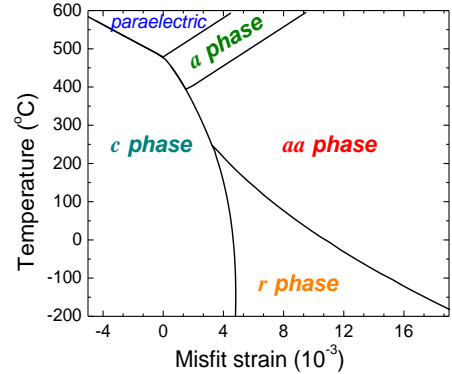

**FIG. 4.** Misfit strain-temperature phase diagram of PbTiO$_3$ multilayer with the volume fraction of passive material $\phi_p = 0.8$.

The phase maps of BaTiO$_3$ multilayers have many similarities with those of PbTiO$_3$ ones, but they are distinguished by the presence of an additional ferroelectric state at $u_m \lesssim 0.8 \cdot 10^{-3}$. Figure 5 shows that the monoclinic *ac* phase ($P_1 \neq 0$, $P_2 = 0$, $P_3 \neq 0$) appears on the ($\phi_p$, $T$) map, which may be attributed to the existence of the orthorhombic phase in BaTiO$_3$ crystals at temperatures between 8°C and −71°C.[33] Remarkably, this phase does not form in epitaxial BaTiO$_3$ films.[5,27] As is seen from Fig. 6, the in-plane and out-of-plane polarization components differ from each other in the multilayer so that the polarization vector in the *ac* phase is not parallel to the face diagonal of the prototypic cubic unit cell.

The threshold volume fraction $\phi_p^0$, above which the multilayer ($u_m$, $T$) diagram becomes qualitatively similar to that of a thin film, equals $\phi_p^0 \cong 0.548$ for the discussed BaTiO$_3$ multilayers. Figure 7 shows the diagram calculated for the BaTiO$_3$ multilayer with $\phi_p = 0.5$. It can be seen that at $\phi_p < \phi_p^0$ the multilayer diagram differs from the phase diagram of BaTiO$_3$ film[5,27] by the presence of the stability ranges of *a* and *ac* phases. Furthermore, the tricritical point exists on the ($\phi_p$, $T$) maps of BaTiO$_3$ multilayers as well. In the range of negative misfit strains, where the transition into the *c* phase takes place, this point is situated at $\phi_p^* \cong 0.554$ in the considered BaTiO$_3$ multilayers. At $u_m > 0$, the critical volume fraction first decreases with $u_m$ and then stabilizes at the value $\phi_p^* \cong 0.238$ in the strain range $u_m \geq 0.228 \times 10^{-3}$, where the tricritical point corresponds to the transition into the *a* phase.

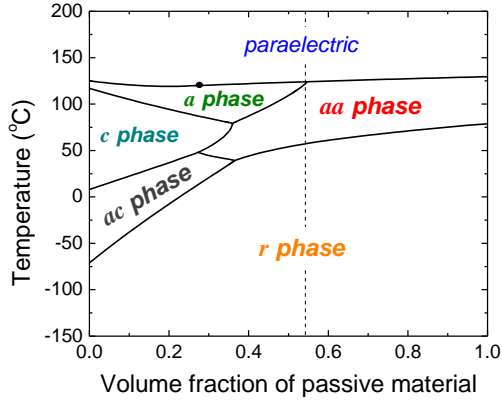

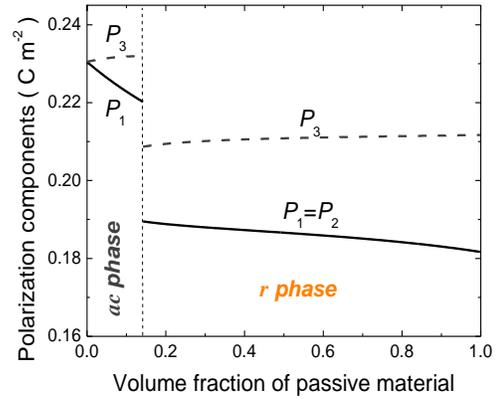

**FIG. 5.** Volume fraction-temperature phase map of BaTiO$_3$ multilayer with the misfit strain $u_m = +0.5 \times 10^{-3}$. The dot shows the tricritical point on the line of ferroelectric phase transition. The dashed line indicates the threshold volume fraction $\phi_p^0$ of passive material, above which the multilayer ($u_m$, $T$) diagram becomes similar to that of the BaTiO$_3$ thin film.

**FIG. 6.** Polarization components $P_i$ in the BaTiO$_3$ multilayer as a function of the volume fraction $\phi_p$ of passive material. The misfit strain in the multilayer is assumed to be $+0.5 \times 10^{-3}$, and the temperature equals −25° C. Note that the difference between the in-plane ($P_1$) and out-of-plane ($P_3$) polarization components in the *ac* phase rapidly increases with the increase of $\phi_p$.

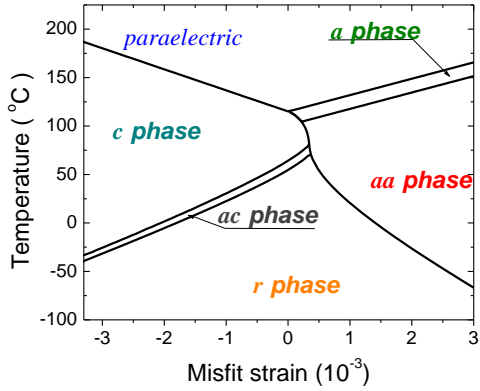

**FIG. 7.** Misfit strain-temperature phase diagram of BaTiO$_3$ multilayer with the volume fraction of passive material $\phi_p = 0.5$. The multilayer phase diagram differs qualitatively from the thin-film diagram by the presence of the stability ranges of *a* and *ac* phases. Note that the volume fraction $\phi_p = 0.5$ is close to the threshold volume fraction $\phi_p^0 \cong 0.548$.

Concluding the discussion of our theoretical results, we would like to make two general remarks. First, in the limit of $\phi_p \to 0$, the *c* and *a* phases become energetically equivalent so that they reduce to the *c* and *a* domain variants of the same tetragonal phase. Hence the transition between these two phase states, which persists in Figs. 3 and 5 down to very small volume fractions of the passive material, disappears at $\phi_p = 0$ in agreement with the behavior of bulk PbTiO$_3$ and BaTiO$_3$ crystals. Second, the formation of the *a* and *ac* phases at $\phi_p < \phi_p^0$ can be explained by the anisotropy $u_{11} \neq u_{22}$ of the total in-plane lattice strains, which is allowed in the multilayers. In contrast, such anisotropy is prohibited in epitaxial thin films in the discussed case of an isotropic biaxial misfit strain $u_{m1} = u_{m2} = u_m$.

## IV. CONCLUSIONS

In this paper, we studied the phase states of free-standing multilayers, where ferroelectric slabs are interleaved by passive layers with linear elastic properties. Our analytical and numerical calculations lead to the following main predictions:

(i) In the limiting cases of $\phi_f \to 0$ and $\phi_f \to 1$, the polarization states of a ferroelectric multilayer coincide with those of a thin film and a bulk crystal, respectively.

(ii) At intermediate volume fractions $0 < \phi_f < 1$, the misfit strain-temperature phase diagram of a ferroelectric multilayer may be very different from that of an epitaxial ferroelectric film. In particular, a new phase forms in BaTiO$_3$ multilayers, which does not appear in BaTiO$_3$ bulk crystals and epitaxial films.

(iii) The paraelectric to ferroelectric phase transformation in PbTiO$_3$ and BaTiO$_3$ multilayers changes from the second-order transition to the first-order one with the increase of the volume fraction $\phi_f$ so that a *tricritical point* appears at a certain value of $\phi_f$.

(iv) The multilayer composition $\phi_f$ may strongly affect the ferroelectric transition temperature at constant misfit strain.

Thus, our theoretical analysis confirmed that free-standing ferroelectric-elastic multilayers demonstrate crossover from thin-film to bulk polarization behavior. We also demonstrated that the ferroelectric transition in such multilayers and their phase states may exhibit important features not displayed by both bulk crystals and thin films. Our results on the polarization states of PbTiO$_3$ and BaTiO$_3$ multilayers provide the basis for the future calculations of their dielectric, pyroelectric, and piezoelectric properties. Importantly, by tuning the volume fraction of ferroelectric material it may be possible to enhance the physical characteristics of a multilayer. Therefore, free-standing ferroelectric multilayers may be advantageous for applications in microelectronic devices.

Moreover, the developed phase diagrams open the way for nonlinear thermodynamic calculations of the strain-mediated direct magnetoelectric effect in ferroelectric-ferromagnetic multilayers, which up to now were performed for the film-substrate hybrids only.[35,36] Our approach may be also applied to the theoretical description of the nanolamellar $BaTiO_3$–$CoFe_2O_4$ bicrystal fabricated recently.[37] Since free-standing ferroelectric multilayers and superlattices have additional tunable parameter and the clamping of active component is reduced here in comparison with conventional epitaxial thin films, these material systems may exhibit enhanced static and dynamic magnetoelectric responses making them attractive for device applications.

## ACKNOWLEDGMENT

The authors like to thank Dr. Vladimir G. Kukhar for valuable suggestions concerning numerical computations of phase diagrams.